\documentclass[twocolumn,floatfix,aps,superscriptaddress,prb]{revtex4}
\usepackage{graphicx}
\usepackage{amsmath}
\usepackage{amssymb}
\usepackage{mathtools}
\usepackage{bm}

\usepackage{color}
 
\begin{document}

\title{Valley switch in a graphene superlattice due to pseudo-Andreev reflection}
\author{C. W. J. Beenakker}
\affiliation{Instituut-Lorentz, Universiteit Leiden, P.O. Box 9506, 2300 RA Leiden, The Netherlands}
\author{N. V. Gnezdilov}
\affiliation{Instituut-Lorentz, Universiteit Leiden, P.O. Box 9506, 2300 RA Leiden, The Netherlands}
\author{E. Dresselhaus}
\affiliation{Instituut-Lorentz, Universiteit Leiden, P.O. Box 9506, 2300 RA Leiden, The Netherlands}
\author{V. P. Ostroukh}
\affiliation{Instituut-Lorentz, Universiteit Leiden, P.O. Box 9506, 2300 RA Leiden, The Netherlands}
\author{Y. Herasymenko}
\affiliation{Instituut-Lorentz, Universiteit Leiden, P.O. Box 9506, 2300 RA Leiden, The Netherlands}
\author{\.{I}. Adagideli}
\affiliation{Faculty of Engineering and Natural Sciences, Sabanci University, Orhanli-Tuzla, Istanbul, Turkey}
\author{J. Tworzyd{\l}o}
\affiliation{Institute of Theoretical Physics, Faculty of Physics, University of Warsaw, ul.\ Pasteura 5, 02--093 Warszawa, Poland}

\date{May 2018}
\begin{abstract}
Dirac electrons in graphene have a valley degree of freedom that is being explored as a carrier of information. In that context of ``valleytronics'' one seeks to coherently manipulate the valley index. Here we show that reflection from a superlattice potential can provide a valley switch: Electrons approaching a pristine-graphene--superlattice-graphene interface near normal incidence are reflected in the opposite valley. We identify the topological origin of this valley switch, by mapping the problem onto that of Andreev reflection from a topological superconductor, with the electron-hole degree of freedom playing the role of the valley index. The valley switch is ideal at a symmetry point of the superlattice potential, but remains close to 100\% in a broad parameter range.
\end{abstract} 
\maketitle

\textit{Introduction ---}
The precession of a spin in a magnetic field has analogues for the pseudospin degrees of freedom that characterize quasiparticles in condensed matter. The K,${\rm K}'$ valley index of Dirac electrons in graphene is a such a pseudospin --- it is actively studied because it might play a role as a carrier of information in ``valleytronics'', the valley-based counterpart of spintronics.\cite{Sch15} The analogue of the magnetic field for valley precession can be provided by a superlattice potential:\cite{Wan15,Xu16,Wu16,Wu17} When graphene is deposited on a substrate with a commensurate honeycomb lattice, the valleys are coupled by the periodic modulation of the potential on the carbon atoms.\cite{Ren15,Gio15,Ven16} The coupling can be represented by an artificial magnetic field\cite{Jua13} that rotates the valley pseudospin as the electron propagates through the superlattice,\cite{Wan15} analogously to the spin-rotation by an exchange field in spintronics.\cite{Han14}

Here we present a valley precession effect without a counterpart in spintronics: A quantized $180^\circ$ precession angle upon reflection, such that the electron switches valleys. This valley switch is analogous to Andreev reflection at the interface between a normal metal and a topological superconductor.\cite{Bee13} Andreev reflection is the reflection as a hole of an electron incident on the superconductor, which happens with unit probability if the normal-superconductor interface contains a Majorana zero-mode. The analogous effect happens in the superlattice because of an anti-unitary symmetry that is formally equivalent to the charge-conjugation symmetry in a superconductor.

\textit{Graphene superlattice with anti-unitary symmetry ---}
We consider the Dirac Hamiltonian\cite{Ren15,Gio15,Ven16}
\begin{equation}
H=\begin{pmatrix}
V_0+\mu&v_{\rm F}p_- &0&\alpha\\
v_{\rm F}p_+&V_0-\mu&-\beta^\ast&0\\
0&-\beta&V_0-\mu&v_{\rm F}p_-\\
\alpha^\ast&0&v_{\rm F}p_+&V_0+\mu
\end{pmatrix}.\label{HDiracdef}
\end{equation}
It acts on the spinor $\Psi=(\psi_{{\rm K}{\rm A}},\psi_{{\rm K}{\rm B}},-\psi_{{\rm K}'{\rm B}},\psi_{{\rm K}'{\rm A}})$ that contains the sublattice (A,B) and valley (K,${\rm K}'$) degrees of freedom of a conduction electron moving in the $x$--$y$ plane of a carbon monolayer (graphene), with velocity $v_{\rm F}$ and momentum $\bm{p}=(p_x,p_y)$, $p_\pm\equiv p_x\pm ip_y$. In terms of Pauli matrices $\sigma_i$ and $\tau_i$ acting, respectively, on the sublattice and valley indices, we may write
\begin{align}
&H=v_{\rm F}(p_x\sigma_x+p_y\sigma_y)+\mu\tau_z\otimes\sigma_z\nonumber\\
&\quad+\tfrac{1}{2}(\tau_x\otimes\sigma_x){\rm Re}\,(\alpha-\beta)-\tfrac{1}{2}(\tau_y\otimes\sigma_y){\rm Re}\,(\alpha+\beta)\nonumber\\
&\quad-\tfrac{1}{2}(\tau_x\otimes\sigma_y){\rm Im}\,(\alpha-\beta)-\tfrac{1}{2}(\tau_y\otimes\sigma_x){\rm Im}\,(\alpha+\beta).
\label{HDiracPauli}
\end{align}
For simplicity, we have shifted the zero of energy such that $V_0=0$.

\begin{figure}[tb]
\centerline{\includegraphics[width=0.9\linewidth]{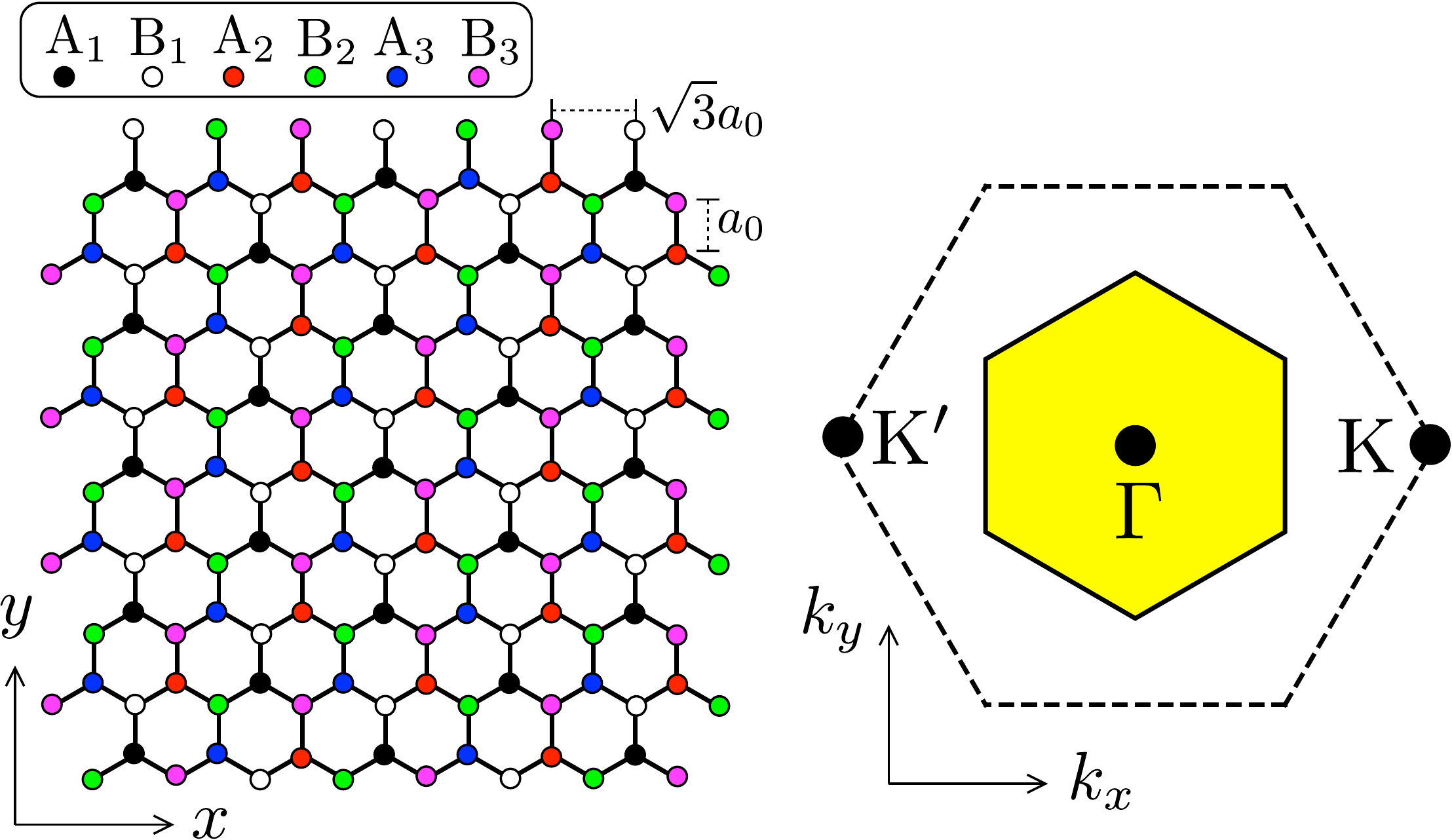}}
\caption{\textit{Left panel:} Hexagonal lattice of graphene, decorated by a periodic potential modulation. Different colors distinguish the carbon atoms on the A and B sublattice, each of which has an ionic potential $V_{{\rm A}_n}$, $V_{{\rm B}_n}$, $n=1,2,3$, induced by the substrate. The lattice constant $a_0$ of the original hexagonal lattice is increased by a factor $\sqrt 3$ in the superlattice. \textit{Right panel:} The Dirac points at the K and ${\rm K}'$ points of the original Brillouin zone of graphene (dashed hexagon) are folded onto the $\Gamma$ point at the center of the superlattice Brillouin zone (yellow hexagon). 
}
\label{fig_lattice}
\end{figure}

An epitaxial substrate induces a periodic potential modulation, which triples the size of the unit cell: it is enlarged by a factor $\sqrt 3\times \sqrt 3$ and contains six rather than two carbon atoms. The parameters $V_0,\mu$ (real) and $\alpha,\beta$ (complex) are determined by the substrate potentials on these six atoms.\cite{note1} The Brillouin zone remains hexagonal, but the two Dirac cones at opposite corners K, ${\rm K}'$ of the original Brillouin zone of graphene are folded onto the center $\Gamma$ of the superlattice Brillouin zone. Depending on the relative magnitude of $\alpha,\beta,\mu$ a gap may open or a linear or quadratic band crossing may appear.\cite{Ren15,Gio15,Ven16}

If there is translational invariance in the $y$-direction,\cite{note2} the momentum component $p_y\equiv q$ is a good quantum number and we may consider the Hamiltonian $H(q)$ at a fixed $q$. Time-reversal symmetry is expressed by
\begin{equation}
(\tau_y\otimes\sigma_y)H^\ast(q)(\tau_y\otimes\sigma_y)=H(-q),\label{TRSeq}
\end{equation}
where the complex conjugation should also be applied to the momentum operator $p_x=-i\hbar\partial/\partial x\equiv-i\hbar\partial_x$. Note the sign change of $q$. An additional anti-unitary symmetry without inversion of $q$ exists if $\beta^\ast=\alpha$,
\begin{align}
&H(q)=v_{\rm F}(-i\hbar\sigma_x\partial_x+q\sigma_y)+\mu\tau_z\otimes\sigma_z\label{Hqalphaisbeta}\\
&\quad-(\tau_y\otimes\sigma_y){\rm Re}\,\alpha-(\tau_x\otimes\sigma_y){\rm Im}\,\alpha,\;\;
{\rm if}\;\;\beta^\ast=\alpha,\nonumber\\
&\Rightarrow \tau_x H^\ast(q)\tau_x=-H(q).\label{Ceq}
\end{align}

\textit{Topological phase transitions ---}
The symmetry \eqref{Ceq} is formally identical to charge-conjugation symmetry in a superconductor, where $\tau_x$ switches electron and hole degrees of freedom. Because the symmetry operation ${\cal C}=\tau_x{\cal K}$ (with ${\cal K}$ = complex conjugation) squares to $+1$, it is symmetry class D in the Altland-Zirnbauer classification of topological states of matter.\cite{Alt97,note3} This correspondence opens up the possibility of a phase transition into a phase that is analogous to a topological superconductor\cite{Has10,Qi11} --- with the K and ${\rm K}'$ valleys playing the role of electron and hole.

The $q$-dependent topological quantum number ${\cal Q}(q)$ of the Hamiltonian \eqref{Hqalphaisbeta} follows from Kitaev's Pfaffian formula,\cite{Kit01}
\begin{equation}
{\cal Q}(q)={\rm sign}\, {\rm Pf}[\tau_x H(q)]_{p_x=0}.\label{QKitaev}
\end{equation}
(The multiplication by $\tau_x$ ensures that the Pfaffian Pf is calculated of an antisymmetric matrix.) We find
\begin{equation}
{\cal Q}(q)={\rm sign}\,(v_{\rm F}^2q^2+\mu^2-|\alpha|^2).\label{Qqresult}
\end{equation}
The graphene superlattice is always topologically trivial (${\cal Q}=+1$) for large $|q|$. However, provided that $|\mu|<|\alpha|$, it is topologically nontrivial (${\cal Q}=-1$) in an interval near $q=0$. There is a pair of topological phase transitions at
\begin{equation}
q=\pm q_{\rm c},\;\;v_{\rm F}q_{\rm c}=\sqrt{|\alpha|^2-\mu^2}.\label{qpmdef}
\end{equation}

To probe the topological phase transition, we contact the graphene superlattice at $x=0$ and $x=L$ by pristine-graphene electrodes, heavily doped so that the Fermi energy in the pristine graphene is high above the Dirac point. By analogy with the conductance of a Kitaev wire,\cite{Bee13} the $2\times 2$ reflection matrix $\bm r$ in the large-$L$ limit should be fully diagonal in the topologically trivial phase and fully off-diagonal in the topologically nontrivial phase. In the superconducting problem this means that there is complete Andreev reflection from the topological superconductor. Here the analogue is a complete \textit{valley switch:} An incident electron in valley K is reflected in the other valley ${\rm K}'$ with unit probability when $|q|<q_{\rm c}$.

\begin{figure}[tb]
\centerline{\includegraphics[width=0.6\linewidth]{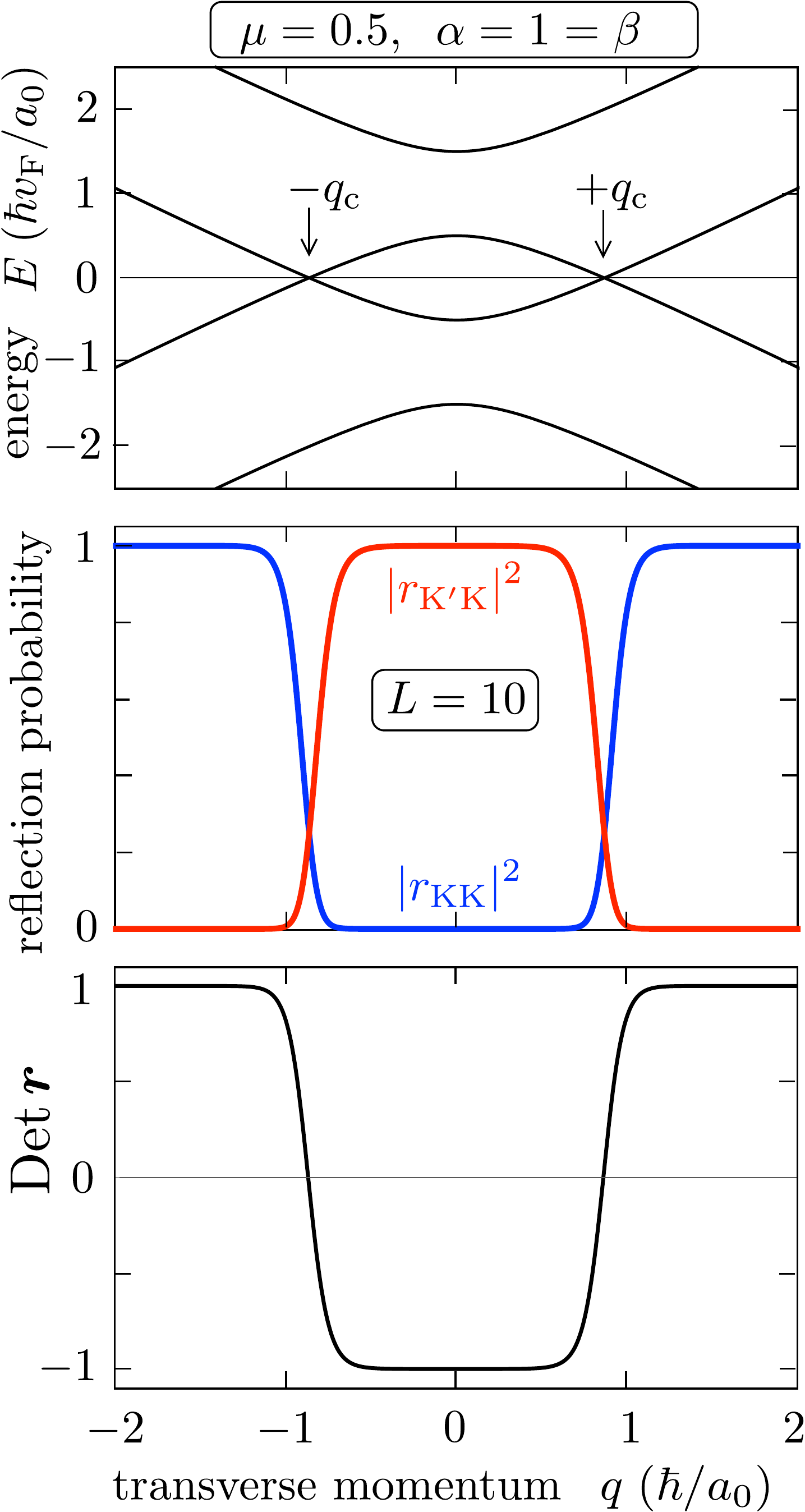}}
\caption{\textit{Top panel:} Energy spectrum of superlattice graphene with Hamiltonian \eqref{Hqalphaisbeta} at $\mu=0.5$, $\alpha=1=\beta$ (in units of $\hbar v_{\rm F}/a_0$). \textit{Middle panel:} Corresponding reflection probabilities from a superlattice-graphene strip of length $L=10\,a_0$ in the $x$-direction, connected at $x=0$ and $x=L$ to heavily doped pristine-graphene electrodes. In the topologically nontrivial regime near normal incidence ($|q|<q_{\rm c}=\sqrt{|\alpha|^2-\mu^2}$) the reflection is completely into the opposite valley. \textit{Bottom panel:} Topological order parameter ${\cal O}_L(q)={\rm Det}\,\bm{r}$, computed from Eq.\ \eqref{Detr}. The switch from intra-valley to inter-valley reflection lines up with the switch from ${\cal O}_L(q)=+1$ to $-1$.
}
\label{fig_reflectionswitch}
\end{figure}

\textit{Valley switch ---}
Let us see how this expectation is borne out by an explicit calculation (detailed in the appendix). The energy spectrum of the Hamiltonian \eqref{Hqalphaisbeta} is shown in Fig.\ \ref{fig_reflectionswitch}. The intervalley coupling splits the Dirac cone at the $\Gamma$ point into a pair of cones at $\bm{k}=(0,\pm q_{\rm c})$ which are gapped out if $|\mu|>|\alpha|$. The same figure shows the $E=0$ reflection probabilities $|r_{\rm{K}'{\rm K}}|^2$ and $|r_{\rm{KK}}|^2$ with and without a valley switch. We clearly see the transition from complete inter-valley to complete intra-valley reflection at $q=\pm q_{\rm c}$.

The analytical formulas that govern this transition are simplest for the special case $\mu=0$, when $q_{\rm c}=|\alpha|/v_{\rm F}$. We find the reflection matrix
\begin{align}
&{\bm r}=\begin{pmatrix}
r_{\rm{KK}}&r_{\rm{KK}'}\\
r_{\rm{K}'{\rm K}}&r_{\rm{K}'\rm{K'}}
\end{pmatrix},\;\;
r_{\rm{KK}}=r_{\rm{K}'\rm{K'}}=-Z^{-1}\sinh 2qL,\nonumber\\
&r_{\rm{KK}'}^{\vphantom{\ast}}=r_{\rm{K}'{\rm K}}^\ast=-iZ^{-1}(\alpha/|\alpha|)\sinh 2q_{\rm c}L,\label{rmuis0}\\
&Z=\cosh 2q_{\rm c}L+\cosh 2qL.\nonumber
\end{align}
(We have set $\hbar\equiv 1$.) In the topologically trivial regime $|q|>q_{\rm c}$ the off-diagonal elements of $\bm r$ vanish $\propto \exp[-2(|q|-q_{\rm c})L]$, while in the topologically nontrivial regime $|q|<q_{\rm c}$ it is the diagonal elements that vanish $\propto \exp[-2(q_{\rm c}-|q|)L]$. 

To confirm that the valley switch at $|q|=q_{\rm c}$ is due to a topological phase transition in a \textit{finite} system, we calculate the $L$-dependent topological order parameter \cite{Akh11}
\begin{equation}
\begin{split}
{\cal O}_L(q)&={\rm Det}\,\bm{r}=-\tanh(\xi_- L)\tanh(\xi_+ L),\\
&\xi_\pm=|\alpha|\pm\sqrt{v_{\rm F}^2q^2+\mu^2}.
\end{split}
\label{Detr}
\end{equation}
Unlike the Pfaffian invariant \eqref{QKitaev}, the determinant \eqref{Detr} crosses over smoothly from $+1$ to $-1$ in an interval around $|q|=q_{\rm c}$. This interval becomes narrower and narrower with increasing $L$, approaching the discontinuous topological phase transition in the infinite-$L$ limit. At $|q|=q_{\rm c}$ the determinant of $\bm r$ vanishes, signifying the opening of a reflectionless mode, a mode that is transmitted with unit probability through the superlattice.

\begin{figure}[tb]
\centerline{\includegraphics[width=0.6\linewidth]{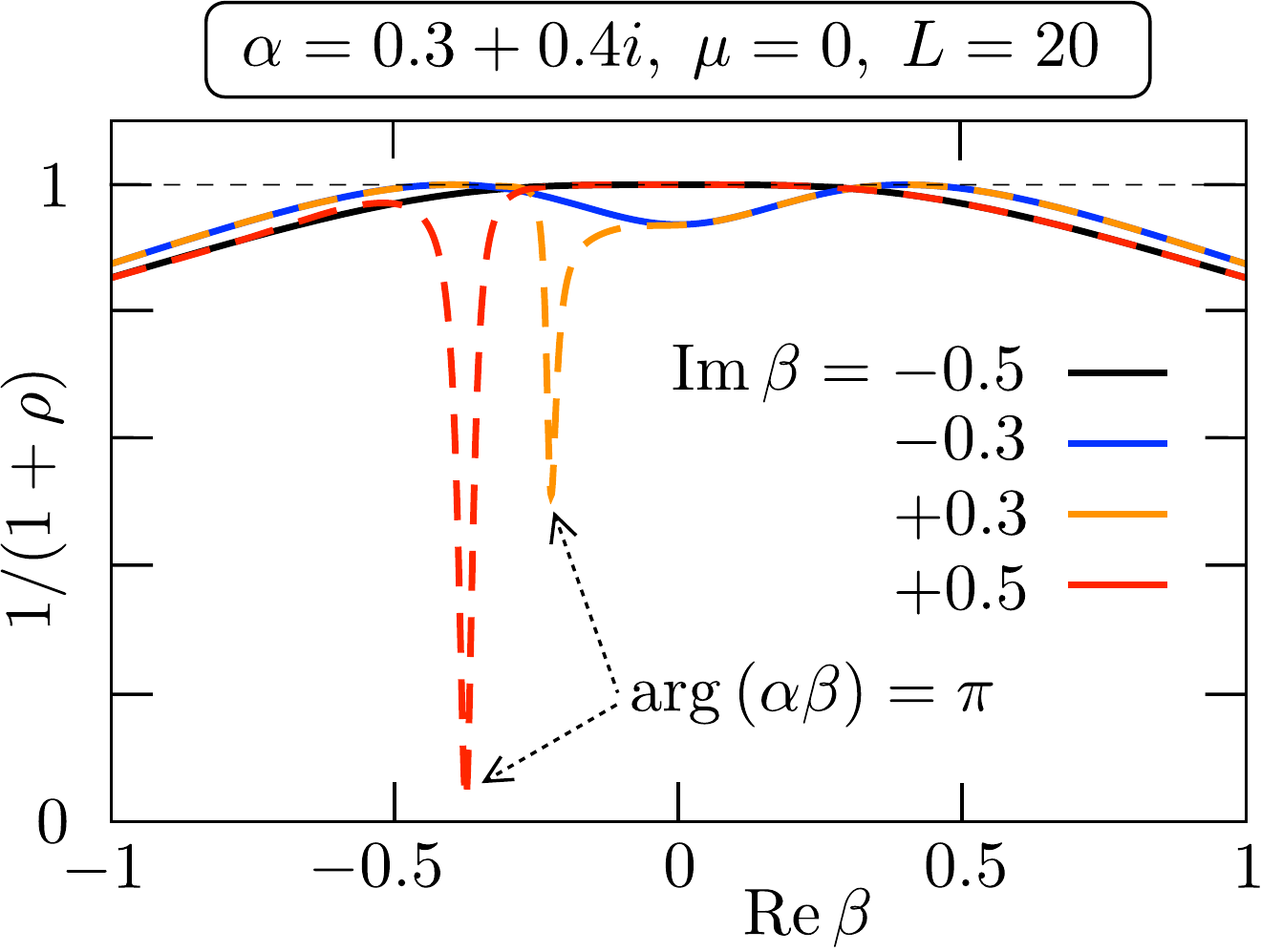}}
\caption{Plot of the fraction $1/(1+\rho)=|r_{\rm{K}'{\rm K}}|^2/(|r_{\rm KK}|^2+|r_{\rm{K}'{\rm K}}|^2)$ of the reflected intensity at normal incidence ($q=0$) that is reflected in the opposite valley, calculated from Eq.\ \eqref{rhoresult}. The reflection with valley switch happens at close to unit probability provided that the product $\alpha\beta$ stays away from the negative real axis.
}
\label{fig_robustness2}
\end{figure}

\begin{figure}[tb]
\centerline{\includegraphics[width=0.6\linewidth]{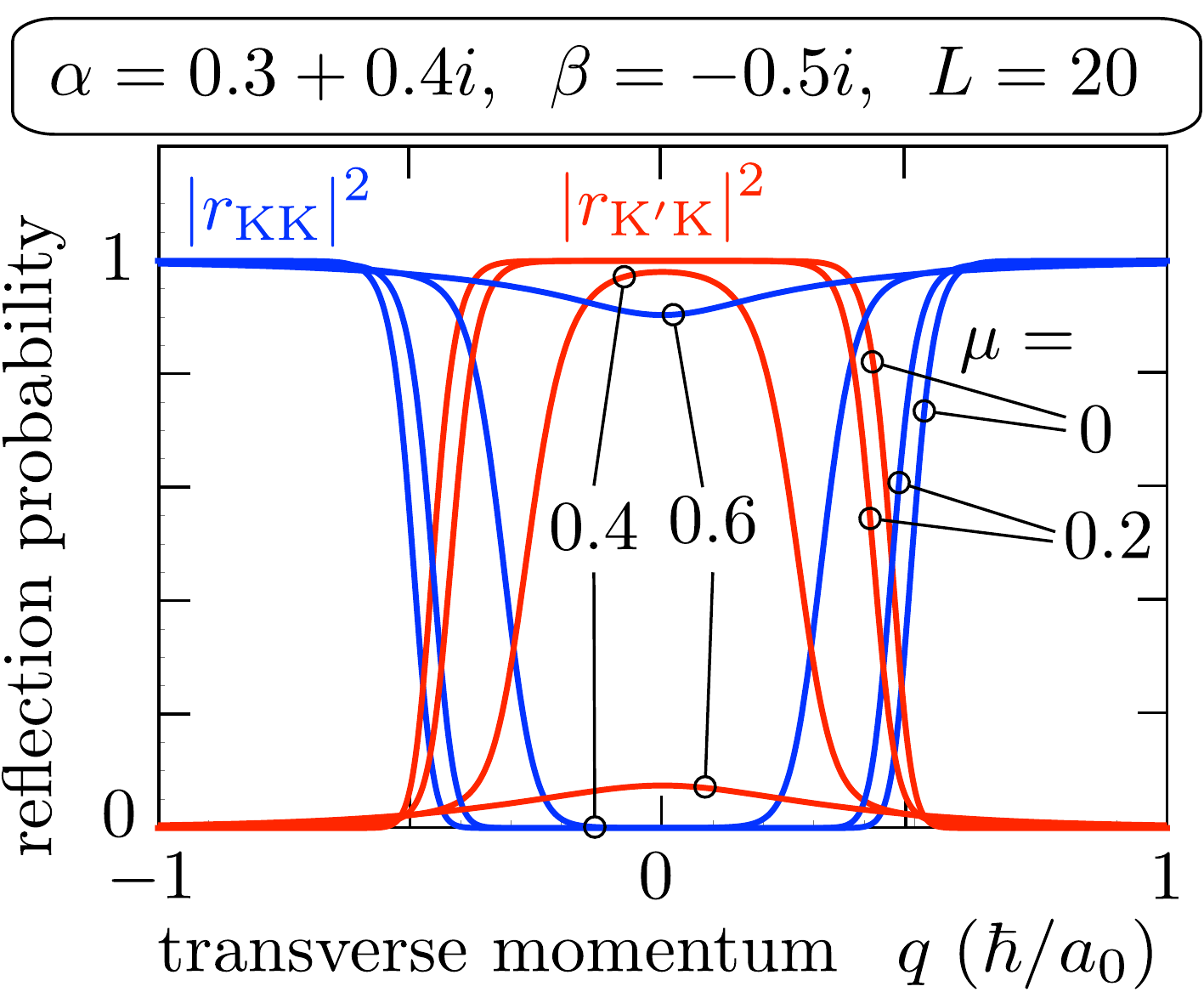}}
\caption{Reflection probabilities from the superlattice-graphene strip with and without a valley switch, for four values of $\mu$. The parameters $\alpha,\beta$ are quite far from the ideal symmetry point $\beta^\ast=\alpha$, but still the reflection near normal incidence happens predominantly in the opposite valley ($|r_{\rm{K}'{\rm K}}|^2\approx 1$) provided $\mu<|\alpha|\approx|\beta|$.
}
\label{fig_robustness}
\end{figure}

\textit{Robustness of the valley switch ---}
The anti-unitary symmetry \eqref{Ceq} is broken if we move away from the symmetry point $\beta^\ast=\alpha$. Let us find out how sensitive the valley switch effect is to the symmetry breaking. The simplest formulas appear for $\mu=0$. To quantify the magnitude of the valley switch we calculate the ratio $\rho=|r_{\rm KK}/r_{\rm{K}'{\rm K}}|^2$ at $q=0$. We find
\begin{align}
\rho&=\frac{(|\alpha|^2-|\beta|^2)^2}{4|\alpha^\ast \Xi+\beta\Xi^\ast|^{2}},
\;\;\Xi=\sqrt{\alpha\beta}\coth(\sqrt{\alpha\beta}L),\label{rhoresult}\\
&\rightarrow \tfrac{1}{4}(|\alpha|-|\beta|)^2\,|\alpha\beta|^{-1}\;\;\text{for}\;\;L\rightarrow\infty\;\;\text{if}\;\;\text{arg}\,(\alpha\beta)\neq\pi,\nonumber
\end{align}
see Fig.\ \ref{fig_robustness2} for a plot. Only the absolute value of the complex amplitudes $\alpha,\beta$ enters in the large-$L$ limit, provided that $\alpha\beta$ is not on the negative real axis (when reflection is quenched by the opening of a propagating mode).\cite{note4} The valley switch happens with 100\% probability if $|\alpha|=|\beta|$, but the ratio $|\alpha/\beta|$ may differ from unity by as much as a factor of two and still 90\% of the reflected intensity at normal incidence happens with a valley switch. Fig.\ \ref{fig_robustness} shows that this robustness to variation of parameters persists for $\mu\neq 0$ -- as long as $\mu<|\alpha|\approx |\beta|$.

Concerning the robustness of the valley switch to disorder, we firstly note that forward scattering events which only couple the transverse momenta within the topologically nontrivial interval $|q|<q_{\rm c}$ do not spoil the topological protection. Secondly, large-angle scattering with a mean free path longer than the penetration depth $\xi=1/q_{\rm c}$ into the superlattice will also not affect the valley switch.

\textit{Conclusion ---} We have presented a topological mechanism that switches the ${\rm K,K}'$ valley index of Dirac fermions in graphene. Unlike scattering processes that require control on the atomic scale, such as intervalley reflection from an armchair edge, our valley switch relies on the long-range effect of a superlattice potential. 

The valley switch is protected by a topological invariant, the Pfaffian \eqref{QKitaev}, originally introduced by Kitaev to describe a topologically nontrivial superconductor.\cite{Kit01} Because of this topological protection the switch happens with 100\% probability even in the presence of a large Fermi energy mismatch at the interface with the superlattice. It is analogous to the 100\% Andreev reflection from a Majorana zero-mode, which is also unaffected by a Fermi energy mismatch at the interface with the superconductor.\cite{Bee13}

We have identified the symmetry point of the superlattice Hamiltonian at which an anti-unitary symmetry appears that is analogous to charge-conjugation symmetry in a superconductor, and we have checked that the valley switch remains close to 100\% in a broad parameter range around that symmetry point. We expect that the analogy between intervalley reflection and Andreev reflection revealed in this work can provide further useful additions to the valleytronics toolbox.\cite{Sch15}

\textit{Acknowledgements ---}
This research was supported by the Netherlands Organization for Scientific Research (NWO/OCW) and by an ERC Synergy Grant. ED was sponsored by a Fulbright grant.

\appendix

\section{Calculation of the scattering matrix of the graphene superlattice}
\label{app_Smatrix}

The calculation of the scattering matrix of the graphene superlattice region $0<x<L$, sandwiched between heavily doped pristine-graphene contacts, proceeds as follows. (See Refs.\ \onlinecite{S_Two06,S_Sny08} for similar calculations in graphene.)

We start from the Dirac Hamiltonian $H(p_x,p_y)$, given by Eq.\ \eqref{HDiracPauli}. We consider solutions of the Dirac equation $H\Psi=E\Psi$ at energy $E$ that are plane waves in the $y$-direction, $\Psi(x,y)=\Psi_q(x)e^{iqy}$. The four-component spinor $\Psi_q(x)$ in the region $0<x<L$ is a solution of
\begin{equation}
\frac{\partial}{\partial x}\Psi_q(x)=\Xi(q)\Psi_q(x),\;\;
\Xi(q)=i{v}_{\rm F}^{-1}\sigma_x[E- H(0,q)],\label{dpsidx}
\end{equation}
resulting in the transfer matrix
\begin{equation}
\Psi_q(L)={\cal T}(q)\Psi_q(0),\;\;{\cal T}(q)=e^{\Xi(q)}.\label{transfermatrix}
\end{equation}

The next step is to transform to a basis of right-moving and left-moving modes in the contact regions $x<0$, $x>L$. The Dirac Hamiltonian in those regions is
\begin{equation}
H_{\rm contact}=v_{\rm F}(p_x\sigma_x+p_y\sigma_y)-V_{\rm doping}.\label{Hcontact}
\end{equation}
[We use the same valley-isotropic basis $\Psi=(\psi_{{\rm K}{\rm A}},\psi_{{\rm K}{\rm B}},-\psi_{{\rm K}'{\rm B}},\psi_{{\rm K}'{\rm A}})$ as in Eq.\ \eqref{HDiracPauli}.] In the limit $V_{\rm doping}\rightarrow\infty$ of infinitely doped contacts the right-moving modes $\Psi_+(x,y)$ and left-moving modes $\Psi_-(x,y)$ are given for $x<0$ by
\begin{equation}
\begin{split}
&\Psi_+=c_{\rm K}^+ e^{ikx+iqy}\begin{psmallmatrix}
1\\ 1\\ 0 \\ 0
\end{psmallmatrix}+c_{{\rm K}'}^+e^{ikx+iqy}\begin{psmallmatrix}
0\\ 0\\ 1 \\ 1
\end{psmallmatrix},\\
&\Psi_-=c_{\rm K}^-e^{-ikx+iqy}\begin{psmallmatrix}
1\\ -1\\ 0 \\ 0
\end{psmallmatrix}+c_{{\rm K}'}^-e^{-ikx+iqy}\begin{psmallmatrix}
0\\ 0\\ 1 \\ -1
\end{psmallmatrix},
\end{split}
\end{equation}
with $v_{\rm F}k=V_{\rm doping}\rightarrow\infty$. The same expression with $x\mapsto x-L$ applies for $x>L$.

The transfer matrix in the basis $(c_{\rm K}^+,c_{\rm K}^-, c_{{\rm K}'}^+,c_{{\rm K}'}^-)$ is
\begin{equation}
\tilde{\cal T}(q)={\cal H}{\cal T}(q){\cal H},\;\;{\cal H}=\frac{1}{\sqrt{2}}\begin{psmallmatrix}
1&1&0&0\\
1&-1&0&0\\
0&0&1&1\\
0&0&1&-1
\end{psmallmatrix}.\label{Hadamard}
\end{equation}
After this ``Hadamard transform''\cite{S_Sny08} we can directly read off the elements of the reflection matrix $\bm{r}$ from the $x=0$ interface,
\begin{subequations}
\begin{align}
&{\bm r}=\begin{pmatrix}
r_{\rm{KK}}&r_{\rm{KK}'}\\
r_{\rm{K}'{\rm K}}&r_{\rm{K}'\rm{K'}}
\end{pmatrix}=-(\tilde{\cal T}_{--})^{-1}\cdot\tilde{\cal T}_{-+},\\
&\tilde{\cal T}_{--}=\begin{pmatrix}
\tilde{\cal T}_{22}&\tilde{\cal T}_{24}\\
\tilde{\cal T}_{42}&\tilde{\cal T}_{44}
\end{pmatrix},\;\;\tilde{\cal T}_{-+}=\begin{pmatrix}
\tilde{\cal T}_{21}&\tilde{\cal T}_{23}\\
\tilde{\cal T}_{41}&\tilde{\cal T}_{43}
\end{pmatrix}.
\label{rresult}
\end{align}
\end{subequations}
The final results are lengthy and not recorded here, but they are easily derived using a computer algebra system.


\begin{thebibliography}{99}
\bibitem{Sch15} J. R. Schaibley, H. Yu, G. Clark, P. Rivera, J. S. Ross, K. L. Seyler, W. Yao, and X. Xu, \textit{Valleytronics in 2D materials}, Nature Reviews Materials \textbf{1}, 16055 (2016).
\bibitem{Wan15} S. K. Wang and J. Wang, \textit{Valley precession in graphene superlattices}, Phys. Rev. B \textbf{92}, 075419 (2015).
\bibitem{Xu16} Fuming Xu, Zhizhou Yu, Yafei Ren, Bin Wang, Yadong Wei, and Zhenhua Qiao, \textit{Transmission spectra and valley processing of graphene and carbon nanotube superlattices with inter-valley coupling}, New J. Phys. \textbf{18}, 113011 (2016).
\bibitem{Wu16} Xiuqiang Wu and Hao Meng, \textit{Tunable valley filtering in graphene with intervalley coupling}, EPL \textbf{114}, 37008 (2016).
\bibitem{Wu17} Qing-Ping Wu, Zheng-Fang Liu, Ai-Xi Chen, Xian-Bo Xiao, Heng Zhang, and Guo-Xing Miao, \textit{Valley precession and valley polarization in graphene with inter-valley coupling}, J. Phys. Condens. Matter \textbf{29}, 395303 (2017).
\bibitem{Ren15} Yafei Ren, Xinzhou Deng, Zhenhua Qiao, Changsheng Li, Jeil Jung, Changgan Zeng, Zhenyu Zhang, and Qian Niu, \textit{Single-valley engineering in graphene superlattices}, Phys. Rev. B \textbf{91}, 245415 (2015).
\bibitem{Gio15} G. Giovannetti, M. Capone, J. van den Brink, and C. Ortix, \textit{Kekul\'{e} textures, pseudo-spin one Dirac cones and quadratic band crossings in a graphene-hexagonal indium chalcogenide bilayer}, Phys. Rev. B \textbf{91}, 121417(R) (2015).
\bibitem{Ven16} J. W. F. Venderbos, M. Manzardo, D. V. Efremov, J. van den Brink, and C. Ortix, \textit{Engineering interaction-induced topological insulators in a $\sqrt 3\times\sqrt 3$ substrate-induced honeycomb superlattice}, Phys. Rev. B \textbf{93}, 045428 (2016).
\bibitem{Jua13} F. de Juan, \textit{Non-Abelian gauge fields and quadratic band touching in molecular graphene}, Phys. Rev. B \textbf{87}, 125419 (2013).
\bibitem{Han14} W. Han, R. K. Kawakami, M. Gmitra, and J. Fabian, \textit{Graphene spintronics}, Nature Nanotech. \textbf{9}, 794 (2014).
\bibitem{Bee13} C. W. J. Beenakker, \textit{Search for Majorana fermions in superconductors}, Annu. Rev. Con. Mat. Phys. \textbf{4}, 113 (2013).
\bibitem{note1} As derived in Ref.\ \onlinecite{Ven16}, the relationship between the parameters $V_0,\mu,\alpha,\beta$ in the Dirac Hamiltonian \eqref{HDiracdef} and the ionic potentials $V_{{\rm A}_n},V_{{\rm B}_n}$, $n=1,2,3$, in Fig.\ \ref{fig_lattice} is: $6V_0=\sum_n (V_{{\rm A}_n}+V_{{\rm B}_n})$, $6\mu=\sum_n (V_{{\rm A}_n}-V_{{\rm B}_n})$, $6\alpha=2V_{{\rm A}_1}-V_{{\rm A}_2}-V_{{\rm A}_3}+i\sqrt{3}(V_{{\rm A}_2}-V_{{\rm A}_3})$, $6\beta=2V_{{\rm B}_1}-V_{{\rm B}_2}-V_{{\rm B}_3}+i\sqrt{3}(V_{{\rm B}_2}-V_{{\rm B}_3})$. When comparing with Ref.\ \onlinecite{Ven16}, note that we are using a basis in which the single-valley Dirac Hamiltonian $v_{\rm F}(\bm{p}\cdot\bm{\sigma})$ is the same in both valleys.
\bibitem{note2} For definiteness, we are considering the orientation of the graphene lattice shown in Fig.\ \ref{fig_lattice}, with the $x$-direction parallel to the zigzag edge. A rotation of the lattice by an angle $\phi$ is equivalent to multiplication of $\alpha$ and $\beta$ by $e^{i\phi}$.
\bibitem{Alt97} A. Altland and M. R. Zirnbauer, \textit{Novel symmetry classes in mesoscopic normal-superconducting hybrid structures}, Phys. Rev. B \textbf{55}, 1142 (1997).
\bibitem{note3} For $\mu=0$ the Hamiltonian \eqref{Hqalphaisbeta} also has chiral symmetry (it anticommutes with $\sigma_z$), so then the symmetry class would be BDI rather than D. The difference is not essential for our analysis.
\bibitem{Has10} M. Z. Hasan and C. L. Kane, \textit{Topological insulators}, Rev. Mod. Phys. \textbf{82}, 3045 (2010).
\bibitem{Qi11} X.-L. Qi and S.-C. Zhang, \textit{Topological insulators and superconductors}, Rev. Mod. Phys. \textbf{83}, 1057 (2011).
\bibitem{Kit01} A. Kitaev, \textit{Unpaired Majorana fermions in quantum wires}, Phys. Usp. \textbf{44} (suppl.), 131 (2001).
\bibitem{Akh11} A. R. Akhmerov, J. P. Dahlhaus, F. Hassler, M. Wimmer, and C. W. J. Beenakker, \textit{Quantized conductance at the Majorana phase transition in a disordered superconducting wire},  Phys. Rev. Lett. \textbf{106}, 057001 (2011).
\bibitem{note4} In terms of the atomic potentials, the condition $\rm{arg}\,\alpha\beta=\pi$ is equivalent to $V_{{\rm A}_1}(V_{{\rm B}_3}-V_{{\rm B}_2})+V_{{\rm A}_2}(V_{{\rm B}_2}-V_{{\rm B}_1})+V_{{\rm A}_3}(V_{{\rm B}_1}-V_{{\rm B}_3})=0$, $V_{{\rm A}_1}(V_{{\rm B}_2}+V_{{\rm B}_3}-2V_{{\rm B}_1})+V_{{\rm A}_2}(V_{{\rm B}_1}+V_{{\rm B}_2}-2V_{{\rm B}_3})+V_{{\rm A}_3}(V_{{\rm B}_1}+V_{{\rm B}_3}-2V_{{\rm B}_2})>0$.
\bibitem{S_Two06}  J. Tworzyd{\l}o, B. Trauzettel, M. Titov, A. Rycerz, and C. W. J. Beenakker, \textit{Sub-Poissonian shot noise in graphene}, Phys. Rev. Lett. \textbf{96}, 246802 (2006).
\bibitem{S_Sny08} I. Snyman, J. Tworzyd{\l}o, and C. W. J. Beenakker, \textit{Calculation of the conductance of a graphene sheet using the Chalker-Coddington network model}, Phys. Rev. B \textbf{78}, 045118 (2008).
\end{thebibliography}
\end{document}